# Theoretical calculation of transport properties of oxide material using narrow band model

Hirofumi Kakemoto[1*]

[1]*Clean Energy Research Center, University of Yamanashi, 4-3-11, Takeda, Kofu, Yamanashi 400-8511, Japan*

We report about the results of theoretical calculations of temperature dependence of resistivity ($\rho$) and Seebeck coefficient ($S$) for thermoelectric (TE) and superconductivity (SC) phases by arithmetic equations based on narrow band model with oxygen deficient structure, as the functions of band-filling degree ($F$), and band width ratio of electron and spin states ($W_\sigma/W_D$). The phase diagrams of TE and SC states, and boundary were imaged to the properties of $\rho$ and $S$ as a function of $F$ and $W_\sigma/W_D$.

## 1. Introduction

Nowadays, various oxide materials are reported about their variety of transport properties: super semiconducting (SC), metallic, semiconducting, and thermoelectric (TE). Recently, oxide TE materials are showing the favorable non-dimensional figure of merit ($ZT$). Particularly, $Na_xCoO_2$ (NCO) shows $p$-type TE property, and reported $ZT$=0.8 at 800K [1]. In experimental study, measurement of Seebeck coefficient ($S$) is widely used not only intensity of $S$ but also polarity at low and/or high temperatures, because $S$ can be directly decided a carrier type of sample. In addition, $S$ is highly sensitive for band structure. In theoretical transport calculation, electrical conductivity ($\sigma=1/\rho$), thermal conductivity ($\kappa$), and $S$ of various materials are investigated by Fermi integral method [2], Boltzmann equation using the results of band calculation [3].

In hopping dynamics of $d$ electron in oxide TE material, Heikes formula can also represent $S$ $(=(k_B/q)\ln\{F/(1-F)\}$, i.e. $S$ is similar with "entropy" of thermal flow in crystal) with hopping carriers caused by oxygen vacancy in oxide TE material.[4] In addition, Gasumyants eq. is expanded for Heikes formula, and it is possible to show temperate dependence of $S$ property by using "narrow band model" with oxygen vacancy. Figure 1 shows schematic of band structure (electronic and spin density of state: $D(E)$, $\sigma(E)$, and their band width: $W_D$, $W_\sigma$). The correlation between electron and spin affects to $W_\sigma/W_D$.[5] In this paper, calculation results about the transport property of TE, SC phases, and boundary are reported.

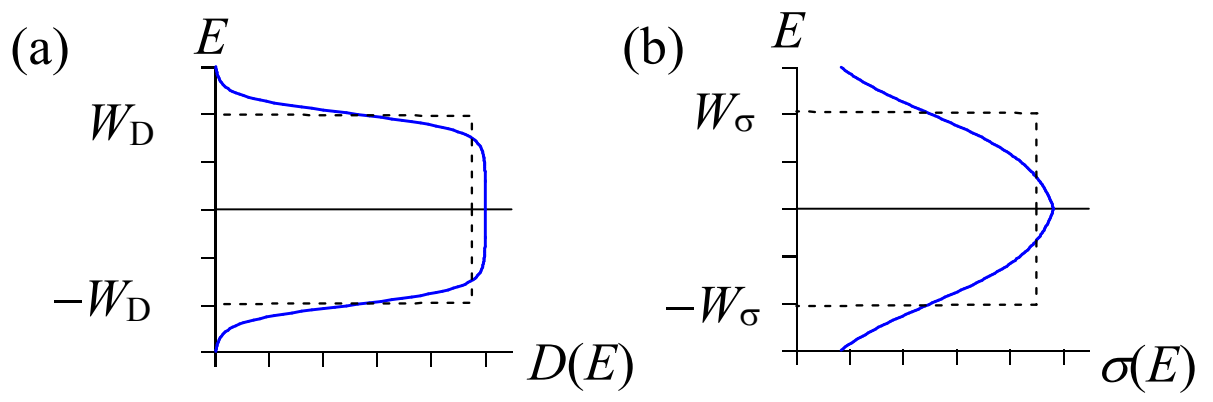

FIG.1 The density of state (solid line), and band width (dashed line) for (a) electron: $D(E)$, $W_D$ and (b) spin: $\sigma(E)$, $W_\sigma$, defined as $W_i=[3\int E^2\chi_i(E)dE/\int\chi_i(E)dE]^{1/2}$.

## 2. Calculation

Figure 2 shows schematics of band structure [6] with oxygen deficient structure for SC [7] and TE phase. The calculations were carried out for SC and TE materials with oxygen deficient.

Temperature dependence of resistivity ($\rho$) and Seebeck coefficient ($S$) were represented by band-filling degree ($F$) and band width ratio of electron and spin states ($C=W_\sigma/W_D$), as fllows, [5]

$$\rho=(1/\sigma)=[1+e^{-2\mu^*}+2e^{-\mu^*}\cosh(C/T^*)]/[e^{-\mu^*}\sinh(C/T^*)] \quad (1),$$

$$S=-(k_B/q)\{(C/T^*)/\sinh(C/T^*))\cdot[e^{-\mu^*}+\cosh(C/T^*)-(T^*/C)(\cosh(\mu^*)+\cosh(C/T^*))\cdot\ln\{(e^{\mu^*}+e^{W_\sigma^*})/(e^{\mu^*}+e^{-W_\sigma^*})\}]-\mu^*\} \quad (2),$$

where $\mu^*=\ln\{\sinh[FW_D^*]/\sinh[(1-F)W_D^*]\}$, $W_\sigma^*=W_\sigma/2k_BT$, $W_D^*=W_D/2k_BT$, $T^*=k_BT/W_D$, and $C=W_\sigma/W_D$.

The equation (2) can be approximated to be $S\sim(-\pi^2/3)(k_B^2T/q)\{d\ln\sigma(E)/dE\}_{E=\mu}$, $T\sim0$ (Sommerfeld approximation), and $S\sim(1/qT)(\mu-\langle E\rangle)$ for high $T$.[8]

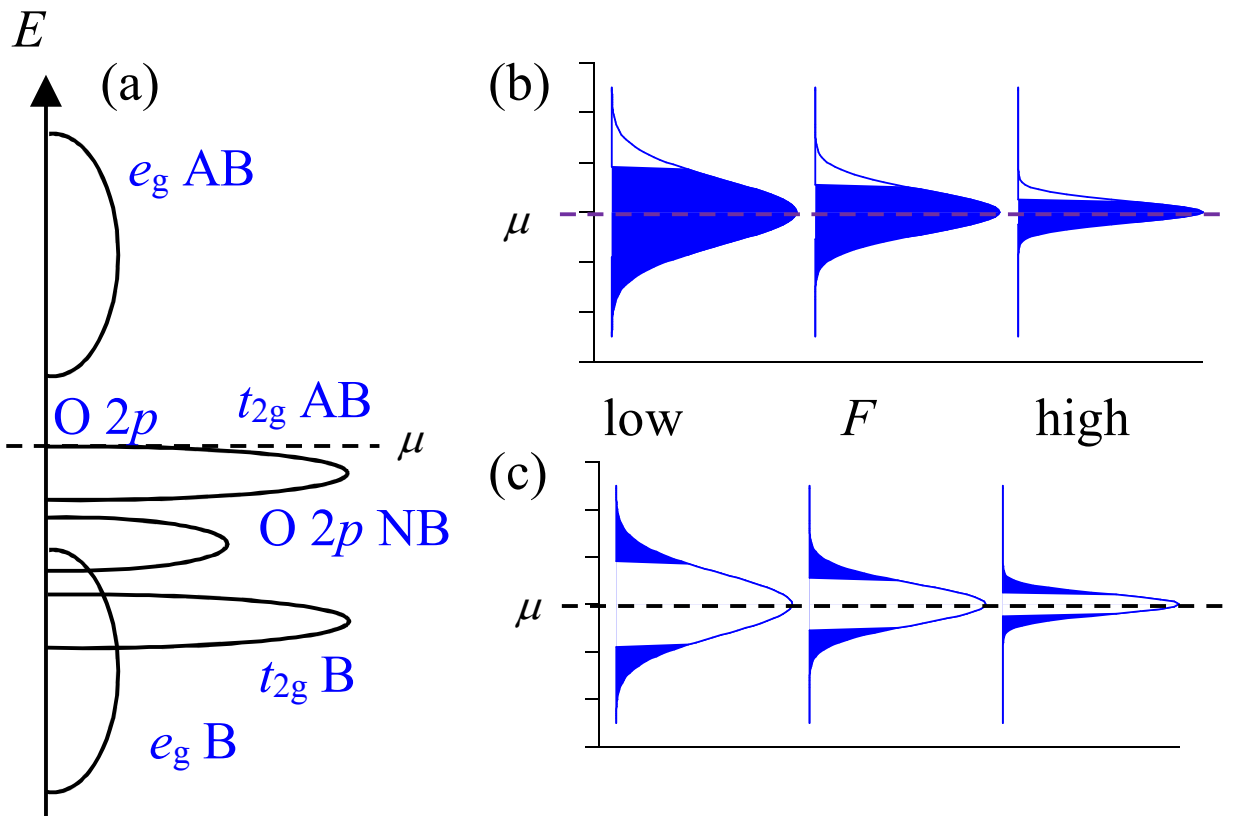

FIG.2 The narrow band model, (a) total band structure, vicinity of Fermi level for oxygen 2$p$ states with vacancy, (b) TE, and (c) SC phases as a function of $F$.

## 3. Results and discussion

Figure 3 shows temperature dependence of resistivity ($\rho$) and Seebeck coefficient ($S$) calculated by narrow band model ($W_\sigma/W_D$=0.5-0.56, $F$=0.5–0.8) using eq.(1-2) for TE phase. In Fig.3(a), as a function of $F$, $\rho$ shows metallic ($F$ < 0.78), and thermally activated features ($F \geq 0.78$). In Fig.3(b), temperature dependence of $S$ shows nonlinear properties. In low $T$, follows as Sommerfeld approximation: $W^*$>>1, $S$~$T$. In high $T$, follows as $W^*$<<1, $S$~$(1/qT)(\mu-\langle E\rangle)$, affected by charge and spin correlation $W_\sigma/W_D$.[5,8] Kaurav et al reported about temperature dependence of $S$ of $Na_xCoO_2$ (NCO), and fitting parameters ($W_\sigma/W_D$=0.56–0.78, $F$=0.56–0.74). The calculated $S$ in Fig.3(b) is hence agreed with TE phase. [9]

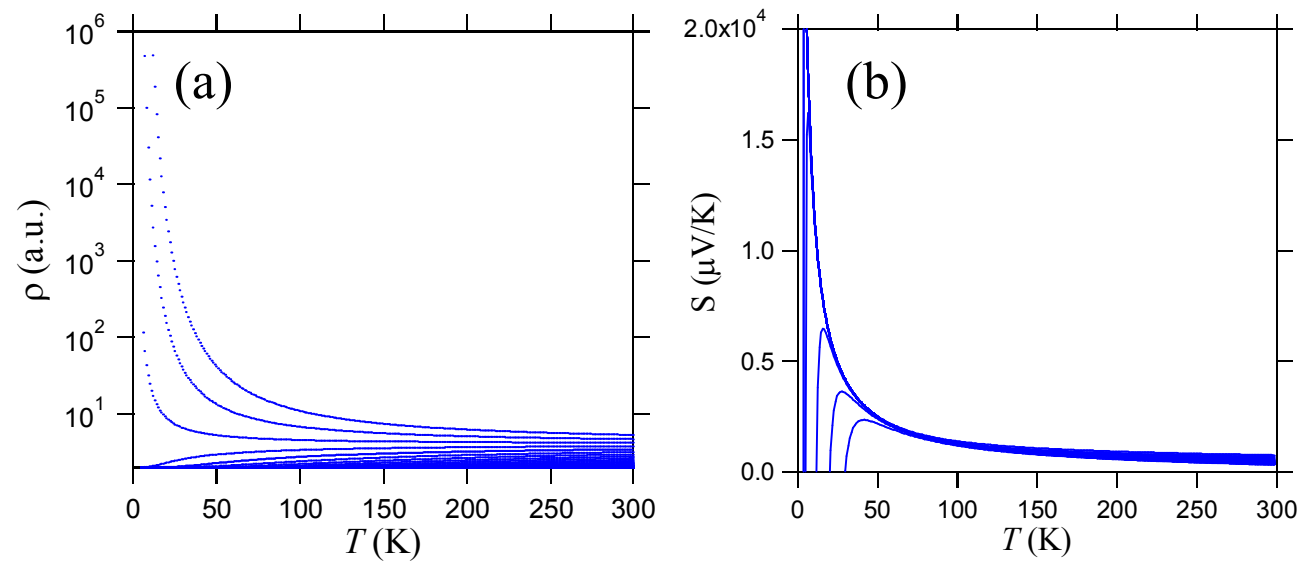

FIG.3 Temperature dependence of (a) resistivity, (b) Seebeck coefficient for TE phase calculated by using eq.(1,2), as the functions of $F$, and $W_\sigma/W_D$.

Figure 4 shows temperature dependence of $\rho$ and $S$ calculated by narrow band model ($W_\sigma/W_D$=0.63eV, $F$=0.5–0.8) using eq.(1-2) for SC state. In Fig.4(a), $\rho$ is decreased with decreasing $T$ as a function of $F$. $S$ becomes zero below Curie temperature ($T_c$) of SC state, as shown in Fig.4(b). Gsumyants et al. reported about $YBa_2Cu_3O_{7-\delta}$, affected by charge and spin correlation $W_\sigma/W_D$=0.3-0.9. [5]

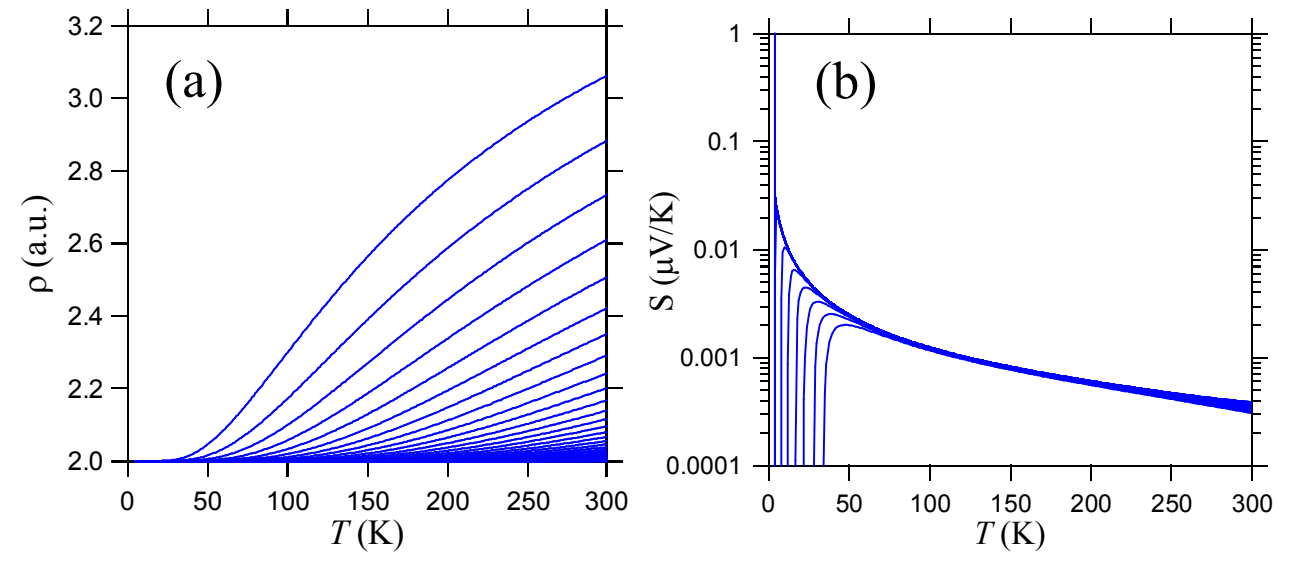

FIG.4 Temperature dependence of (a) resistivity, (b) Seebeck coefficient for SC phase calculated by using eq.(1-2), as the functions of $F$ and $W_\sigma/W_D$.

Figure 5 shows phase diagrams of TE and SC phases appeared in $\rho$ and $S$ versus $T$ properties, calculated as the functions of $F$ (0.5–0.8) and $W_\sigma/W_D$ (0.1–0.7). As shown in Figs.3 and 4, TE and SC phases are also reproduced in Fig.5. In this study, by the narrow band model with oxygen vacancy, the boundary between TE and SC phases can be also visualized about $\rho$ in Fig.5(a) $F$=0.65, (c) $W_\sigma/W_D$=0.2, $S$ in (b) $F$=0.7, and (d) $W_\sigma/W_D$=0.22.

As shown in Fig.2, in narrow band model, 1) $F$ parameter is affected by oxygen vacancy. 2) The band, and band widths are shifting and enlarging by structural transition, A-site vacancy and/or substitution, and intercalation for oxide material.

For example, the vacancies of sodium and oxygen in oxide TE material: NCO are reported, and it affects in TE properties.[10] In addition, NCO is modified to be $Na_xCoO_2\cdot1.3H_2O$, and it shows SC state at $T_c$=4K.

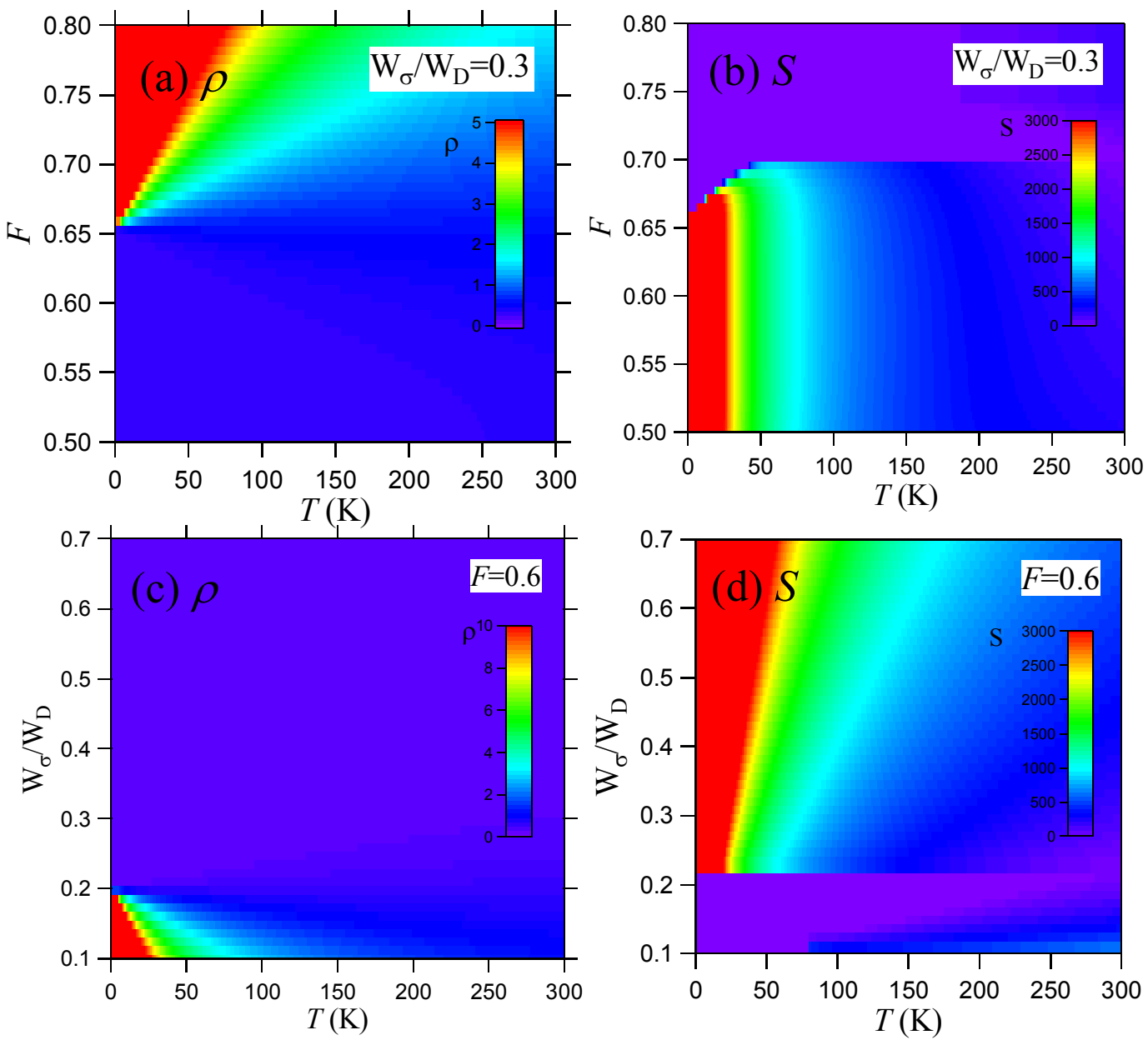

FIG.5 The phase diagrams of (a) resistivity and (b) Seebeck coefficient as a function of $F$, and (c) resistivity and (d) Seebeck coefficient as a function of $W_\sigma/W_D$.

Table I Calculating conditions: parameters in eq.(1,2). [5,9]

| Material | Type | $F$ | $W_D$ (eV) | $W_\sigma$ (eV) | $W_\sigma/W_D$ | Condition |
|---|---|---|---|---|---|---|
| TE : $Na_xCoO_{2-\delta}$ | $p$ | 0.5–0.8 | 0.1–1.0 | 0.3 | 0.5–0.56, 0.1–0.7 | I |
| SC : $YBa_2Cu_3O_{7-\delta}$ | $p$ | 0.5–0.8 | 0.1–1.0 | 0.3 | 0.5–0.63, 0.1–0.7 | II |

The strong two dimensional property of the layered Cu oxides plays an important roles in SC state. Similarly, the separation of the $CoO_2$ layers by introducing $H_2O$ molecules would be important for SC state.[11]

It is revealed that $F$ affected oxygen vacancy and band width ratio: $W_\sigma/W_D$ for $CoO_2$ layers are useful parameters, respectively. It would be understood about TE and SC properties and phase boundary of various oxide materials by using narrow band model.

## 4. Conclusion

Temperature dependences of resistivity ($\rho$) and Seebeck coefficient ($S$) of superconducting (SC) and thermoelectric (TE) phases were calculated by using narrow band model with oxygen deficient. The detailed parameters (band-filling degree: $F$, band widths: $W_D$, $W_\sigma$) were used for calculations by using arithmetic equation.

The calculated carve showed nonlinear $S$ at low $T$: $S$~$T$, Sommerfeld approximation, and at high $T$: $S$~$(1/eT)(\mu-<E>)$. $\rho$ and $S$ versus $T$, and boundary of SC and TE phases were clearly visualized as the functions of $F$ and $W_\sigma/W_D$. The narrow band model is hence useful for simply understanding about transport properties of TE and SC phase and their boundary, and the parameters ($F$, $W_D$, and $W_\sigma$) are possible to be estimated in experimental studies.

In future study, the investigation of $n$-type Nb related TE oxide will be carried out using narrow band model.

Acknowledgment

This work was partly supported by Japan Society for the Promotion of Science (JSPS) KAKENHI Grant-in-Aid for Scientific Research(C) Number JP25410238.

*Present address: 1-15-11, Sakura-cho, Tsuchiura, Ibaraki, 300-0037, Japan, Techno Pro R&D company (Tsukuba branch), Techno Pro Inc.

e mail: hkakemoto@yamanashi.ac.jp